\documentclass[twocolumn,pre,showpacs]{revtex4}
\usepackage{amssymb}
\usepackage{amsfonts}
\usepackage{amsmath}
\usepackage{graphicx}

\begin{document}

\title{Thermodynamics of quantum jump trajectories in systems driven by
classical fluctuations}
\author{Adri\'{a}n A. Budini}
\affiliation{Consejo Nacional de Investigaciones Cient\'{\i}ficas y T\'{e}cnicas, Centro
At\'{o}mico Bariloche, Av. E. Bustillo Km 9.5, (8400) Bariloche, Argentina}
\date{\today}

\begin{abstract}
The large-deviation method can be used to study the measurement trajectories
of open quantum systems. For optical arrangements this formalism allows to
describe the long time properties of the (non-equilibrium) photon counting
statistics in the context of a (equilibrium) thermodynamic approach defined
in terms of dynamical phases and transitions between them in the trajectory
space [J.P. Garrahan and I. Lesanovsky, Phys. Rev. Lett. \textbf{104},
160601 (2010)]. In this paper, we study the thermodynamic approach for
fluorescent systems coupled to complex reservoirs that induce stochastic
fluctuations in their dynamical parameters. In a fast modulation limit the
thermodynamics corresponds to that of a Markovian two-level system. In a
slow modulation limit, the thermodynamic properties are equivalent to those
of a finite system that in an infinite-size limit is characterized by a
first-order transition. The dynamical phases correspond to different
intensity regimes, while the size of the system is measured by the
transition rate of the bath fluctuations. As a function of a dimensionless
intensive variable, the first and second derivative of the thermodynamic
potential develop an abrupt change and a narrow peak respectively. Their
scaling properties are consistent with a double-Gaussian probability
distribution of the associated extensive variable.
\end{abstract}

\maketitle

\section{Introduction}

The interaction of small quantum systems with an infinite set of
uncontrollable degrees of freedom leads to non-equilibrium time irreversible
evolutions. This is the main topic of the theory of open quantum systems 
\cite{breuerbook}. In this context, a diverse kind of physical systems can
be described through a quantum master equation, which defines the dynamics
of their density matrix operators.

Quantum optical arrangements fall in the previous category \cite%
{carmichaelbook,plenio,barchielli,optica,optica1}. The irreversible dynamics
is induced by the background electromagnetic field, which leads to the
natural radiative decay of the system. The opposite mechanism, where the
system becomes excited to higher energy levels, can be induced by the
interaction with an external laser field. The competition between both
effects produces a continuous emission of photons.

The detection of the radiated photons occurs at random times. Successive
measurement realizations provide an ensemble of trajectories, which define a
stochastic (point) process \cite{vanKampen}. Its statistics can be described
through different approaches, such as generating operator techniques \cite%
{cook}, or the quantum jump approach \cite%
{breuerbook,barchielli,carmichaelbook,plenio}. These formalisms can also be
utilized in the description of fluorescent systems coupled to classically
fluctuating reservoirs \cite{barkaiChem,michel,jung}. The interaction with
the bath is modeled through a set of random processes that modify the
characteristic parameters of the system \cite%
{zhengBloch,heSpectral,BarkaiExactMandel,brownGeneratriz,slow}. In these
cases, the main task is to relate the environment fluctuations with the
photon-counting statistics \cite%
{barkaiChem,michel,jung,zhengBloch,heSpectral,BarkaiExactMandel,brownGeneratriz,slow,SMS,SMSJumps,budini,osadko}%
.

In contrast\ to the previous approaches, statistical mechanics provides the
theoretical tools for describing systems in thermal equilibrium \cite{raquel}%
. It allows to get the dependence of extensive thermodynamic observables as
a function of the conjugate (intensive) variables. Phase transition points
are defined by nonanalycities of a thermodynamic potential in the parameter
space.

While both non-equilibrium (small) quantum systems and thermodynamic ones
are described with intrinsically different approaches, in a recent
contribution Garrahan and Lesanovsky related both kind of descriptions \cite%
{garrahan}. The bridge between both areas is provided by the large-deviation
(LD) theory \cite{touchette}. This formalism is concerned with the
exponential decay of probabilities corresponding to large fluctuations in a
stochastic system. It allows to describe its ensemble of trajectories in the
same way as equilibrium statistical mechanics describes ensemble of
configurations in phase space \cite{touchette,lecomte,vander,sollich,evans}.
The role of the thermodynamic variables is played by dynamical order
parameters or their associated conjugate fields. The existence of
\textquotedblleft space-time\textquotedblright\ phase transitions in glassy
systems \cite{glassy} was established through this approach.

We notice that the dynamic and physical properties, such as subPoissonian
photon statistic, photon-antibunching, spectrum peaks (Mollow triplet),
squeezed noise, bistability, etc., that can be found in quantum optical
systems are very well known \cite{optica,optica1}. The breakthrough
introduced in Ref. \cite{garrahan} is not directly related with those
phenomena. The main idea was to apply the LD formalism to the measurement
trajectories of simple open quantum systems, such as (two) three-level
fluorescent systems and a micromaser. The extra physical aspect that can be
analyzed with the LD approach is the asymptotic (long time) statistical
properties of the measurement trajectories, such as for example the
statistic of the number of detected photons in the case of fluorescent
systems or the number of atoms leaving the cavity in a given state for the
micromaser. The LD approach, by going beyond the central limit theorem \cite%
{touchette}, allows to describe the asymptotic regime with a set of
functions that in a statistical mechanics interpretation play the role of
entropy and free energy. By modifying the system parameters and a conjugate
dimensionless (dynamical) order parameter, properties such as scale
invariance points, crossover between distinct dynamical phases, and an
actual first-order phase transition were found in the thermodynamic approach 
\cite{garrahan}. As these properties are related with physical observables
defined from the ensemble of measurement trajectories they should be
detectable in experiments.

The main goal of this paper is to extend the analysis of Ref. \cite{garrahan}
to the case of fluorescent systems coupled to complex self-fluctuating
reservoirs. The system-bath dynamics is described through a density matrix
formalism \cite{SMS,SMSJumps}. Our interest is to characterize the
thermodynamic formalism associated to the photon counting probabilities as a
function of the statistical properties of the environment fluctuations. We
show that in a fast modulation limit, i.e., when the characteristic time of
the bath fluctuations is the minor time scale of the problem, the
thermodynamic frame reduces to that of a Markovian fluorescent two-level
system. A scale invariance point is recovered for a special set of system
parameter values. On the other hand, in a slow modulation limit the
thermodynamical properties of the measurement trajectories are equivalent to
those of finite systems that in an infinite-size limit develop a first-order
phase transition \cite{binderBook,binder,landauIsing,challa}. Here, the
phases are related with different intensity regimes of the scattered
radiation, while the (thermodynamic) size of the system is associated to the
characteristic rate of the bath fluctuations. The finite-size effects are
similar to those found in the Ising or q-state Potts models \cite%
{landauIsing,challa}. These properties are shown through the thermodynamical
response functions, i.e., the first and second derivatives of the
thermodynamic potential with respect to an intensive parameter.

The paper is outlined as follows. In Sec. II, based on the theoretical
results established in Ref. \cite{garrahan}, we define the thermodynamic
approach for an arbitrary counting process. In Sec. III we define the
density matrix evolution and photon counting statistics of the system of
interest, i.e., a fluorescent (two-level) system driven by classical
fluctuations. In Sec. IV we analyze the thermodynamic approach in the limit
of fast environment fluctuations, while the case of slow fluctuations is
developed in Sec. V. The conclusions are presented in Sec. VI.

\section{Thermodynamics of counting processes}

Here, we define the thermodynamic formalism \cite{garrahan} for an arbitrary
counting process \cite{vanKampen}. In the next section, it is build up from
the photon statistics of a fluorescent system driven by classical
fluctuations.

A counting process is defined by a set of trajectories, each one consisting
in a series of consecutive events occurring at random times \cite{vanKampen}%
. It can \ be statistically characterized by a set of probabilities $%
\{P_{n}(t)\}_{n=0}^{\infty },$ satisfying $0\leq P_{n}(t)\leq 1,$ and the
normalization%
\begin{equation}
\sum_{n=0}^{\infty }P_{n}(t)=1.
\end{equation}%
Each $P_{n}(t)$ is the probability of occurrence of $n$-events up to time $%
t. $ From these objects, we introduce an associated stochastic process
defined by the probabilities%
\begin{equation}
q_{n}(t)\equiv \frac{1}{Z_{t}(s)}P_{n}(t)e^{-sn},  \label{q}
\end{equation}%
where $s$ is a real parameter. Consistently with the condition $%
\sum_{n=0}^{\infty }q_{n}(t)=1,$ the function $Z_{t}(s)$\ is defined by%
\begin{equation}
Z_{t}(s)\equiv \sum_{n=0}^{\infty }P_{n}(t)e^{-sn}.  \label{partition}
\end{equation}%
Hence, $Z_{t}(s)$ is the generating function \cite{vanKampen} of the
original counting process,%
\begin{equation}
Z_{t}(s)=\langle \langle \exp [-sn_{st}(t)]\rangle \rangle _{\{P\}}.
\label{Z_Paverage}
\end{equation}%
Here, $n_{st}(t)$ is the (stochastic) number of events up to time $t$ in a
given realization while $\langle \langle \cdots \rangle \rangle _{\{P\}}$
denotes an average over the realizations associated to the set $%
\{P_{n}(t)\}_{n=0}^{\infty }.$

From the transformation (\ref{q}), one can deduce that unlikely events of
the counting process $\{P_{n}(t)\}_{n=0}^{\infty }$ becomes typical events
in the $s$-ensemble \cite{touchette}, i.e., in the set of realizations
defined by the probabilities $\{q_{n}(t)\}_{n=0}^{\infty }.$ At each time $%
t, $ the rare events has associated a thermodynamic-like structure. The
consistency of this affirmation becomes evident after introducing the
corresponding statistical objects. A thermodynamic entropy function $S_{t}$
can be defined as the Shannon entropy of the $s$-ensemble, 
\begin{equation}
S_{t}\equiv -\sum_{n=0}^{\infty }q_{n}(t)\log [q_{n}(t)].  \label{Entropy}
\end{equation}%
By reading the function $Z_{t}(s)$ as a partition function \cite{raquel}, we
define a \textquotedblleft free energy function\textquotedblright\ or
\textquotedblleft grand (thermodynamic) potential\textquotedblright\ as%
\begin{equation}
\Theta _{t}\equiv -\log [Z_{t}(s)].
\end{equation}%
A \textquotedblleft internal energy\textquotedblright\ is defined as%
\begin{equation}
\left\langle \left\langle E\right\rangle \right\rangle _{t}\equiv
-\sum_{n=0}^{\infty }q_{n}(t)\log [P_{n}(t)],  \label{InternalEnergy}
\end{equation}%
while the \textquotedblleft average particle number\textquotedblright\ reads%
\begin{equation}
\left\langle \left\langle N\right\rangle \right\rangle _{t}\equiv
\sum_{n=0}^{\infty }q_{n}(t)n.
\end{equation}%
Then, it is straightforward to relate the previous objects through the\
thermodynamic relation%
\begin{equation}
\Theta _{t}=\left\langle \left\langle E\right\rangle \right\rangle
_{t}-S_{t}+s\langle \langle N\rangle \rangle _{t}.  \label{Grand}
\end{equation}%
In fact, in units of energy where $kT=1,$ $k$ denoting the Boltzmann
constant and $T$ temperature, and defining a (dimensionless)
\textquotedblleft chemical potential\textquotedblright\ $\mu \equiv -s,$ the
previous relation arises in the description of thermodynamical (equilibrium)
processes carried out in open systems that can exchange both heat and matter
with their surroundings \cite{raquel}. Therefore, both the energy and
particle number can fluctuate. Consistently with a statistical derivation
based on maximizing entropy, $\left\langle \left\langle E\right\rangle
\right\rangle _{t}$ and $\langle \langle N\rangle \rangle _{t}$ can be read
as the constraints on the average energy and particle number respectively.
Here, the average, denoted as $\left\langle \left\langle \cdots
\right\rangle \right\rangle _{t},$ is defined by the set of probabilities $%
\{q_{n}(t)\}_{n=0}^{\infty }.$

The thermodynamic interpretation allows us to write the average number
(extensive variable) as the derivative of the (pseudo) grand potential $%
\Theta _{t}$ with respect to the (pseudo) chemical potential $s$ \cite%
{raquel} (intensive variable), 
\begin{equation}
\langle \langle N\rangle \rangle _{t}=\frac{\partial }{\partial s}\Theta
_{t},  \label{Average}
\end{equation}%
while the average of the centered quadratic fluctuations follows from the
second derivative,%
\begin{equation}
\langle \langle \Delta N^{2}\rangle \rangle _{t}\equiv \langle \langle
N^{2}\rangle \rangle _{t}-\langle \langle N\rangle \rangle _{t}^{2}=-\frac{%
\partial ^{2}}{\partial s^{2}}\Theta _{t}.  \label{Variance}
\end{equation}%
These (two) thermodynamic relations can alternatively be derived by writing $%
\langle \langle N\rangle \rangle _{t}$ and $\langle \langle \Delta
N^{2}\rangle \rangle _{t}$ as the average and variance of the number of
events up to time $t$ associated to the set of probabilities $%
\{q_{n}(t)\}_{n=0}^{\infty }.$

The thermodynamic frame [Eq. (\ref{Grand})] is parametrized by the time $t.$
In a long time regime, for ergodic processes, it is expected that all
averaged quantities (strictly, all cumulants) become proportional to the
evaluation time $t.$ Hence, the normalized asymptotic average values%
\begin{equation}
\langle \langle \cdots \rangle \rangle \equiv \lim_{t\rightarrow \infty }%
\frac{1}{t}\langle \langle \cdots \rangle \rangle _{t},  \label{asimptotico}
\end{equation}%
become time independent. In \ this regime, the partition function acquires a
LD form \cite{garrahan}%
\begin{equation}
\lim_{t\rightarrow \infty }Z_{t}(s)\approx \exp [-t\Theta (s)].
\label{Zasimptotico}
\end{equation}%
Then, the previous relations [Eqs. (\ref{Average}) and (\ref{Variance})]
maintain their validity after replacing $\Theta _{t}\rightarrow \Theta (s)$
and $\langle \langle \cdots \rangle \rangle _{t}\rightarrow \langle \langle
\cdots \rangle \rangle .$ Notice that $\Theta (s)$ and all normalized
average values $\langle \langle \cdots \rangle \rangle $ have units of
[1/time].

The pseudo-thermodynamic structure defined previously is associated to the
probabilities (\ref{q}). The LD formalism allows to relate it with an
observable property of the original counting process, i.e., with the long
time behavior of the probabilities $\{P_{n}(t)\}_{n=0}^{\infty }.$
Consistently with Eq. (\ref{Zasimptotico}), their asymptotic structure is
written as $\lim_{t\rightarrow \infty }P_{n}(t)\approx \exp [-t\varphi (%
\frac{n}{t})].$ Taking into account that the LD function $\varphi (N)$ ($%
N=n/t$) also defines the asymptotic behavior of the internal energy (\ref%
{InternalEnergy}) and entropy (\ref{Entropy}), it can be related to the
grand potential through a Legendre-Fenchel transformation \cite%
{garrahan,touchette}, $\varphi (N)=\max_{s}[\Theta (s)-sN],$ which in turn
guarantees that $\Theta (s)$ has convexity properties consistent with the
thermodynamical interpretation.

From the previous relations, it becomes evident that the thermodynamical
potential $\Theta (s)$ provide an alternative and complete characterization
of the asymptotic properties of the set $\{P_{n}(t)\}_{n=0}^{\infty }.$ In
consequence, possible phase transitions happening in the thermodynamic frame
must to be related to strong modifications in the statistical properties of
the original counting process. These relations and its associated
theoretical frame provide an alternative and novel approach for analyzing
measurement trajectories of single open quantum systems subjected to a
continuous measurement process \cite{garrahan}.

\section{Fluorescent systems driven by classical fluctuations}

In many nanoscopic optical systems, such as those analyzed in the context of
single-molecule spectroscopy \cite{barkaiChem,michel,jung}, the randomness
of the photon emission process arises from both the interaction with the
background electromagnetic field and intrinsic environment fluctuations.
This last effect can be modeled by a set of noises that modify (modulate)
the parameters of the system evolution \cite%
{zhengBloch,heSpectral,BarkaiExactMandel,brownGeneratriz,slow}. As
demonstrated in Refs. \cite{SMS,SMSJumps}, the noises modeling can be
reformulated through a density matrix formalism. Here, we present a short
derivation of both the underlying density matrix evolution and the photon
counting statistics.

The environment is defined by a set of (configurational) macrostates, each
one leading to a different system dynamics. The transitions between the bath
states is described by a classical rate equation \cite{vanKampen}%
\begin{equation}
\dfrac{dP_{R}(t)}{dt}=\sum\limits_{R^{\prime }}\phi _{RR^{\prime
}}P_{R^{\prime }}(t)-\sum\limits_{R^{\prime }}\phi _{R^{\prime }R}P_{R}(t),
\label{Classical}
\end{equation}%
where $P_{R}(t)$ is the probability of finding the environment in a given
state $R=1,\cdots R_{\max },$ at time $t.$ The set $\{\phi _{RR^{\prime }}\}$
define the hopping rates. The system density matrix $\rho (t)$\ is described
by a set of auxiliary states $\{\rho _{R}(t)\},$ each one representing the
system dynamic for each bath state. Then, by writing 
\begin{equation}
\rho (t)=\sum\limits_{R}\rho _{R}(t),  \label{Rho_S}
\end{equation}%
and demanding the condition%
\begin{equation}
\mathrm{Tr}[\rho _{R}(t)]=P_{R}(t),
\end{equation}%
where $\mathrm{Tr}[\cdots ]$ denotes a trace operation in the system Hilbert
space, we introduce the evolution \cite{SMS}%
\begin{eqnarray}
\dfrac{d\rho _{R}(t)}{dt}\! &=&\!\dfrac{-i}{\hbar }[H_{R},\rho
_{R}(t)]+\gamma _{R}\mathcal{L}[\rho _{R}(t)]  \label{LindbladRate} \\
&&\!+\sum\limits_{R^{\prime }}\phi _{RR^{\prime }}\rho _{R^{\prime
}}(t)\!-\!\sum\limits_{R^{\prime }}\phi _{R^{\prime }R}\rho _{R}(t).  \notag
\end{eqnarray}%
The last line of this equation introduce a coupling between the auxiliary
states and takes into account the environment fluctuations. The constant $%
\gamma _{R}$ define the natural decay of the system associated to each $R$%
-bath state. Consistently, the Lindblad superoperator $\mathcal{L}[\bullet ]$
reads \cite{breuerbook}%
\begin{equation}
\mathcal{L}[\bullet ]=-\frac{1}{2}\{\sigma ^{\dag }\sigma ,\bullet
\}_{+}+\sigma \bullet \sigma ^{\dag },
\end{equation}%
where $\{\cdots \}_{+}$ denotes an anticommutation operation. The operator $%
\sigma (\sigma ^{\dag })$ is the lowering (raising) operator between the
system eigenstates. It is modeled through a two-level optical transition
with eigenstates $\{|\pm \rangle \}.$ Then, $\sigma =|-\rangle \left\langle
+\right\vert $ and $\sigma ^{\dag }=|+\rangle \left\langle -\right\vert .$

In the first line of Eq. (\ref{LindbladRate}), $H_{R}$ is the system
Hamiltonian associated to each bath state. It reads \cite{SMS,SMSJumps}%
\begin{equation}
H_{R}=\frac{\hbar \omega _{R}}{2}\sigma _{z}+\frac{\hbar \Omega _{R}}{2}%
(\sigma ^{\dagger }e^{-i\omega _{L}t}+\sigma e^{+i\omega _{L}t}),
\label{Laser}
\end{equation}%
where $\sigma _{z}$ is the $z$-Pauli matrix in the basis $\{|\pm \rangle \},$
and $\omega _{R}=\omega _{0}+\delta \omega _{R}.$ Therefore, $\omega _{0}$
defines the bare transition frequency of the system, while $\delta \omega
_{R}$ are the spectral shifts induced by the interaction with the bath. The
second contribution takes into account the interaction with the external
laser field. $\Omega _{R}$ is the effective Rabi frequency associated to
each reservoir state, while $\omega _{L}$ is the frequency of the external
laser excitation.

The sum structure Eq. (\ref{Rho_S}), added to the local character of the
evolution (\ref{LindbladRate}), imply the presence of strong non-Markovian
effects in the system dynamics \cite{SMS}. On the other hand, Eq. (\ref%
{LindbladRate}) does not take into account light assisted processes. While
these phenomena appear in real experimental situations \cite%
{barkaiChem,michel,jung}, most of the results developed in the next sections
can be easily extended to include such kind of effects \cite{SMS,SMSJumps}.

\subsection*{Photon-counting statistics}

The photon detection statistics can be obtained by expressing the system
density matrix (\ref{Rho_S}) as%
\begin{equation}
\rho (t)=\sum\nolimits_{n=0}^{\infty }\rho ^{(n)}(t).
\end{equation}%
Each state $\rho ^{(n)}(t)$ corresponds to the system state conditioned to $%
n-$photons detection events \cite{plenio,carmichaelbook}. The probability of
counting $n$-photons up to time $t$ reads%
\begin{equation}
P_{n}(t)=\mathrm{Tr}[\rho ^{(n)}(t)].  \label{Pn}
\end{equation}%
This set of probabilities can be obtained through a generating function
approach \cite{vanKampen}. Due to the quantum nature of the system, a
\textquotedblleft generating operator\textquotedblright\ is introduced%
\begin{equation}
\mathcal{G}(t,s)\equiv \sum\nolimits_{n=0}^{\infty }e^{-sn}\rho ^{(n)}(t),
\label{Gdefinicion}
\end{equation}%
where the extra real parameter $s$ plays the same role as in Eq. (\ref{q}).
The partition function Eq. (\ref{partition}) follow as%
\begin{equation}
Z_{t}(s)=\mathrm{Tr}[\mathcal{G}(t,s)].  \label{ZfromG}
\end{equation}%
From the equality $\rho (t)=\mathcal{G}(t,s)|_{s=0},$ the condition $%
Z_{t}(s)|_{s=0}=1$ is consistently satisfied.

The conditional states $\rho ^{(n)}(t)$ can be decomposed into the
contributions associated to each configurational state of the reservoir,
leading to the expression%
\begin{equation}
\mathcal{G}(t,s)=\sum\nolimits_{n=0}^{\infty }e^{-sn}\sum\nolimits_{R}\rho
_{R}^{(n)}(t)\equiv \sum\nolimits_{R}\mathcal{G}_{R}(t,s).
\label{GeneratorOperator}
\end{equation}%
Each matrix $\rho _{R}^{(n)}(t)$ defines the state of the system under the
condition that at time $t$ $n-$photon detection events happened, and the
environment is in the configurational state $R.$ Consistently, each
contribution $\mathcal{G}_{R}(t,s)$ defines the (conditional) generating
operator \textquotedblleft \textit{given}\textquotedblright\ that the
reservoir is in the $R$-state. Its evolution reads \cite{SMS}%
\begin{eqnarray}
\dfrac{d\mathcal{G}_{R}(t,s)}{dt} &=&\dfrac{-i}{\hbar }[H_{R},\mathcal{G}%
_{R}(t,s)]+\gamma _{R}\mathcal{L}_{s}[\mathcal{G}_{R}(t,s)]  \label{G_R} \\
&&+\sum\limits_{R^{\prime }}\phi _{RR^{\prime }}\mathcal{G}_{R^{\prime
}}(t,s)-\sum\limits_{R^{\prime }}\phi _{R^{\prime }R}\mathcal{G}_{R}(t,s), 
\notag
\end{eqnarray}%
where the superoperator $\mathcal{L}_{s}$ is given by%
\begin{equation}
\mathcal{L}_{s}[\bullet ]=-\frac{1}{2}\{\sigma ^{\dag }\sigma ,\bullet
\}_{+}+e^{-s}\ \sigma \bullet \sigma ^{\dag }.  \label{LindbladS}
\end{equation}%
The evolution (\ref{G_R}) can be solved in a Laplace domain. From Eq. (\ref%
{GeneratorOperator}), the partition function (\ref{ZfromG}) can always be
written as a quotient of two polynomial functions, $Z_{u}(s)=f(u)/g(u),$
where $u$ is the Laplace variable. Then, by using the residues theorem, the
grand potential $\Theta (s)$\ [Eq. (\ref{Zasimptotico})] follows from the
larger root of the equation $g(u)=0.$ This property allows us to get all
objects characterizing the $s$-ensemble. In the following section we study
its associated statistic in a fast and slow modulation limits.

In Ref. \cite{garrahan} it was demonstrated that it is always possible to
find a density matrix $\tilde{\rho}(t)$ whose associated measurement
statistics is given by the probabilities $q_{n}(t),$ Eq. (\ref{q}). In the
present context, one can affirm that the $s$-ensemble statistics can be
recovered from a set of auxiliary states $\{\tilde{\rho}_{R}(t)\}_{R=1}^{R_{%
\max }}$ defined as%
\begin{equation}
\tilde{\rho}_{R}(t)=\frac{l_{R}^{1/2}\rho _{R}(t)l_{R}^{1/2}}{\sum_{R}%
\mathrm{Tr}[l_{R}\rho _{R}(t)]},  \label{RhoTilde}
\end{equation}%
where the set of operators $\{l_{R}\}$\ is the left \textquotedblleft
eigenoperator\textquotedblright\ of the evolution (\ref{G_R}).

\section{Fast modulation limit}

The thermodynamic frame associated to Eq. (\ref{ZfromG}) in general cannot
be characterized in an analytical way. Nevertheless, in the limit of fast
and slow environment fluctuations the problem becomes analytically treatable.

For the original ensemble of trajectories [Eq. (\ref{Pn})], the fast
modulation limit refers to the case in which the environment transitions are
much faster than the photon emission process. From Eq. (\ref{LindbladRate}),
this condition can explicitly be written as%
\begin{equation}
\{\phi _{RR^{\prime }}\}\gg \{I_{R}\}.  \label{fast}
\end{equation}%
The constant $I_{R}$ is the intensity associated to the $R$-bath state
[diagonal contribution in (\ref{LindbladRate})], i.e., the intensity of a
Markovian two-level system with natural decay $\gamma _{R},$ Rabi frequency $%
\Omega _{R},$ and detuning $\delta _{R}$ \cite{SMS,SMSJumps},%
\begin{equation}
I_{R}=\frac{\gamma _{R}\Omega ^{2}}{\gamma _{R}^{2}+2\Omega ^{2}+4\delta
_{R}^{2}},  \label{I_R}
\end{equation}%
where $\delta _{R}\equiv \omega _{L}-\omega _{R}.$ Both, $\omega _{L}$ and $%
\omega _{R}$ are defined from Eq. (\ref{Laser}). The inequality (\ref{fast})
implies that the average time between two consecutive photon emissions is
much larger than the characteristic time of the bath fluctuations. Under
this condition, the fluorescent system can be approximated by a Markovian
system whose evolution is defined by the average parameters \cite{slow,SMS}%
\begin{equation}
\!\gamma =\sum_{R}P_{R}^{\infty }\gamma _{R},\ \ \ \omega
=\sum_{R}P_{R}^{\infty }\omega _{R},\ \ \ \Omega =\sum_{R}P_{R}^{\infty
}\Omega _{R}.  \label{averageParameters}
\end{equation}%
The weights $\{P_{R}^{\infty }\}$ are the stationary solution of (\ref%
{Classical}),%
\begin{equation}
P_{R}^{\infty }\equiv \lim_{t\rightarrow \infty }P_{R}(t).
\label{PR_Estacionaria}
\end{equation}%
Therefore, the generating operator [Eq. (\ref{GeneratorOperator})] can be
approximated as%
\begin{equation}
\mathcal{G}(t,s)\simeq \mathcal{G}_{M}(t,s),  \label{GAproximada}
\end{equation}%
where\ $\mathcal{G}_{M}(t,s)$ is defined by the Markovian evolution%
\begin{equation}
\dfrac{d\mathcal{G}_{M}(t,s)}{dt}=-\frac{i}{\hbar }[H,\mathcal{G}%
_{M}(t,s)]+\gamma \mathcal{L}_{s}[\mathcal{G}_{M}(t,s)].  \label{G_Markov}
\end{equation}%
Here, $\mathcal{L}_{s}[\bullet ]$ follows from Eq. (\ref{LindbladS}) while $%
H $\ can be read from Eq. (\ref{Laser}) after replacing $\omega
_{R}\rightarrow \omega $ and $\Omega _{R}\rightarrow \Omega .$ The sub-index 
$M$ indicates the underlying Markovian approximation.

From Eqs. (\ref{Zasimptotico}) and (\ref{ZfromG}), it is possible to
associate a thermodynamic potential $\Theta _{M}(s)$ to the operator $%
\mathcal{G}_{M}(t,s).$ Its thermodynamics cumulants are%
\begin{equation}
\langle \langle N\rangle \rangle _{M}=\frac{\partial \Theta _{M}(s)}{%
\partial s},\ \ \ \ \ \ \ \langle \langle \Delta N^{2}\rangle \rangle _{M}=-%
\frac{\partial ^{2}\Theta _{M}(s)}{\partial s^{2}}.  \label{CumulanteMarkov}
\end{equation}

The analytical expression for $\Theta _{M}(s)$ that can be obtained from Eq.
(\ref{G_Markov}) correspond to the larger root of a fourth degree
polynomial. When the external laser excitation is in resonance with the
(Markovian) system, i.e., $\omega _{L}=\omega ,$ the polynomial is of third
order. Thus, the expression for $\Theta _{M}(s)$ becomes much more simple.
Under the previous condition, from Eqs. (\ref{Zasimptotico}) and\ (\ref%
{ZfromG}), we get%
\begin{equation}
\Theta _{M}(s)=\frac{\gamma }{2}-\frac{1}{6}f(s)+\frac{4\Omega ^{2}-\gamma
^{2}}{2f(s)}.  \label{GranMarkov}
\end{equation}%
The auxiliary function $f(s),$ after introducing the \textquotedblleft
fugacity\textquotedblright\ $z\equiv \exp (-s)$ \cite{raquel}, reads%
\begin{equation}
f(s)=\left( 54z\gamma \Omega ^{2}+\sqrt{(54z\gamma \Omega
^{2})^{2}+27(4\Omega ^{2}-\gamma ^{2})^{3}}\right) ^{1/3}.
\end{equation}%
Notice that when $\gamma =2\Omega $ it follows $\Theta _{M}=\Omega
(1-e^{-s/3}),$ which recovers the result presented in Ref. \cite{garrahan}.
It is characterized by the scale invariant property $\langle \langle \Delta
N^{2}\rangle \rangle _{M}/\langle \langle N\rangle \rangle _{M}=1/3,$ i.e.,
the normalized fluctuations do not depend on $s.$

In Fig. 1, we characterize the thermodynamic frame associated to Eqs. (\ref%
{LindbladRate}) and (\ref{G_R}) in the fast modulation limit. We assume a
two-dimensional configurational space, i.e., the bath is characterized by
only two states, $R=A,B.$ We take $\omega _{R}=\omega _{0}$ and $\Omega
_{R}=\Omega ,$ i.e., the spectral shifts are null and the system-laser
interaction is independent of the bath states. Hence, the reservoir only
affects the natural decay of the system, $\{\gamma _{R}\}.$ Furthermore, the
laser excitation is assumed to be in resonance with the system, $\omega
_{L}=\omega _{0}.$ Under these conditions, we can approximate the grand
potential as $\Theta (s)\approx \Theta _{M}(s)$ [Eqs. (\ref{GranMarkov})].%
\begin{figure}[htbp]
\includegraphics[bb=44 32 430 620,angle=0,width=8.7cm]{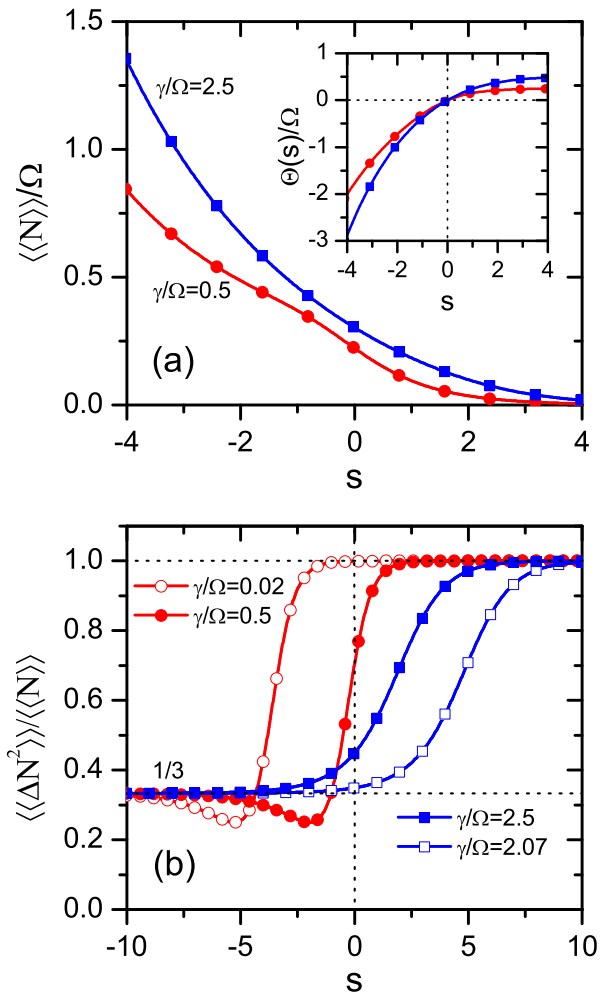}
\caption{(color online) (a) Average number value $\langle \langle N\rangle
\rangle $ as a function of the chemical potential $s$ in the fast modulation
limit. The inset show the respective grand potential $\Theta (s).$ (b) Plot
of the normalized fluctuations $\langle \langle \Delta N^{2}\rangle \rangle
/\langle \langle N\rangle \rangle .$ All curves are almost indistinguishable
with the fitting defined by Eqs. (\protect\ref{CumulanteMarkov}) and (%
\protect\ref{GranMarkov}). The parameters of the Hamiltonian dynamics are $%
\protect\omega _{R}=\protect\omega _{0}$ and $\Omega _{R}=\Omega ,$ while
for the irreversible one read $\protect\gamma _{A}/\Omega =2,$ $\protect%
\gamma _{B}/\Omega =3$ (blue filled squares), and $\protect\gamma %
_{A}/\Omega =0.3,$ $\protect\gamma _{B}/\Omega =0.7$ (red filled circles).
In (b), the parameters of the extra curves are $\protect\gamma _{A}/\Omega
=2.5,$ $\protect\gamma _{B}/\Omega =1.64$ (blue empty squares), and $\protect%
\gamma _{A}/\Omega =0.01,$ $\protect\gamma _{B}/\Omega =0.03$ (red empty
circles). In all cases we take $\protect\phi _{AB}/\Omega =\protect\phi %
_{BA}/\Omega =10.$ The average decay rate, Eq. (\protect\ref%
{averageParameters}), is written in each plot.}
\end{figure}

In Fig. 1(a) we plot the average number $\langle \langle N\rangle \rangle .$
In the inset, we show the corresponding grand thermodynamic potential, which
is obtained as the larger root of an eighth order polynomial function. For
the chosen parameter values, the fast modulation limit is achieved for all
values of $s.$ In fact, in both cases the curves are indistinguishable from
the analytical expressions Eqs. (\ref{GranMarkov}) and (\ref{CumulanteMarkov}%
).  The average rate [Eq. (\ref{averageParameters})] is indicated in the
plots. Notice that for the upper (blue filled squares) curve $\gamma /\Omega
>2,$ while for the lower (red filled circles) curve $\gamma /\Omega <2.$
Each value corresponds respectively to an overdamped [$2<\gamma /\Omega
<\infty $] and underdamped [$0<\gamma /\Omega <2$] regimes of the Markovian
dynamics, Eq. (\ref{G_Markov}).

In Fig. 1(b) we plot the normalized fluctuations $\langle \langle \Delta
N^{2}\rangle \rangle /\langle \langle N\rangle \rangle .$ This function is
almost indistinguishable from the fitting that follows from Eqs. (\ref%
{GranMarkov}) and (\ref{CumulanteMarkov}). In the limit $s\rightarrow
+\infty ,$ asymptotically all curves converge to $1.$ This property is
fulfilled by a Poisson process \cite{vanKampen}. In the limit $s\rightarrow
-\infty ,$ all curves converge to $1/3,$ i.e., the value of the scale
invariant regime \cite{garrahan}. In the curves associated to the red filled
circles and blue filled squares the parameters are the same than in Fig.
1(a). On the other hand, the extra curves associated to the red empty
circles and blue empty squares correspond to a different set of parameter
values that also are in the underdamped and overdamped regimes respectively.
From these curves we deduce that, in both regimes, when $\gamma /\Omega
\rightarrow 2$ the transition between $1/3$ and $1$ occurs for higher values
of the chemical potential, $s\rightarrow +\infty .$ Furthermore, we checked
that when $\gamma /\Omega =2$ a scale invariance property \cite{garrahan} is
recovered, $\langle \langle \Delta N^{2}\rangle \rangle /\langle \langle
N\rangle \rangle =1/3.$ The validity of this behavior occurs for increasing
values of the rates $\phi _{RR^{\prime }}.$

\section{Slow modulation limit}

When $\{\phi _{RR^{\prime }}\}\ll \{I_{R}\},$ the system is able to emits a
huge quantity of photons before a bath transition occurs \cite%
{slow,SMS,SMSJumps}. Therefore, the photon emission process develops a
blinking property, i.e., the intensity randomly changes between the set of
values $\{I_{R}\}.$ Each change in the intensity regime can be associated to
a bath configurational transition. This limit is much more interesting than
the previous one. Signatures of a thermodynamical phase transition can be
found in this regime. First, we describe the emission process through a
stochastic approximation. In a second step, we characterize the
thermodynamic associated to the $s$-ensemble.

\subsection{Stochastic approach}

In the slow modulation limit, the realizations of the photon counting
process [i.e., the realizations associated to the probabilities (\ref{Pn})]
can be approximated by the stochastic process%
\begin{equation}
n_{st}(t)=\sum_{R}\int_{0}^{t}d\tau \delta _{RR_{st}(\tau )}\frac{%
dn_{R}^{st}(\tau )}{d\tau }.  \label{StochasticIntensity}
\end{equation}%
As in Eq. (\ref{Z_Paverage}), $n_{st}(t)$ is the (stochastic) number of
photon detection events up to time $t.$ On the other hand, $R_{st}(\tau )\in
(1,\cdots R_{\max })$ is a random process that indicates which is the state
of the reservoir at time $\tau .$ Then, the contribution\ defined by the
discrete delta function $\delta _{RR_{st}(\tau )}$ does not vanish only when 
$R_{st}(\tau )=R,$ where it is equal to $1.$ Finally, $n_{R}^{st}(\tau )$ is
the (stochastic) number of photon recording events up to time $\tau $
corresponding to a Markovian fluorescent system defined by the decay rate $%
\gamma _{R},$ transition frequency\ $\omega _{R},$ and Rabi frequency $%
\Omega _{R},$ i.e., the parameters associated to each reservoir state.

Consistently with the slow modulation limit, we can assume that $R_{st}(\tau
)$ and the set $\{n_{R}^{st}(\tau )\}$ are statistically independent between
all them. Hence, the average number follows from Eq. (\ref%
{StochasticIntensity}) as%
\begin{equation}
\langle \langle n_{st}(t)\rangle \rangle _{\{P\}}=\sum_{R}\int_{0}^{t}d\tau
\langle \delta _{RR_{st}(\tau )}\rangle \frac{d\overline{n_{R}^{st}}(\tau )}{%
d\tau }.
\end{equation}%
With $\langle \cdots \rangle $ and the overbar $\overline{(\cdots )}$
symbols we denote an average over the realizations of $R_{st}(t)$ and $%
n_{R}^{st}(t)$ respectively. Trivially, one can write $\langle \delta
_{RR_{st}(\tau )}\rangle =P_{R}(\tau ),$ where $\{P_{R}(t)\}$ are the
solution of Eq. (\ref{Classical}). After assuming that $\overline{n_{R}^{st}}%
(0)=0$ and $P_{R}(0)=P_{R}^{\infty }$ [Eq. (\ref{PR_Estacionaria})], we get%
\begin{equation}
\langle \langle n_{st}(t)\rangle \rangle _{\{P\}}=\sum_{R}P_{R}^{\infty }%
\overline{n_{R}^{st}}(t).
\end{equation}%
Therefore, the average number $\langle \langle n_{st}(t)\rangle \rangle
_{\{P\}}$ can be written as a linear combination of the averages $\{%
\overline{n_{R}^{st}}(t)\},$ each one being weighted by the stationary
configurational populations $\{P_{R}^{\infty }\}.$ Taking into account that $%
\overline{n_{R}^{st}}(t)=I_{R}t,$ the normalized asymptotic average value $%
\langle \langle \Delta n_{st}\rangle \rangle _{\{P\}}\equiv
\lim_{t\rightarrow \infty }(1/t)\langle \langle \Delta n_{st}(t)\rangle
\rangle _{\{P\}},$ trivially reads%
\begin{equation}
\langle \langle n_{st}\rangle \rangle _{\{P\}}=\sum_{R}P_{R}^{\infty }I_{R}.
\label{medioStoca}
\end{equation}

The second cumulant, defined by%
\begin{equation}
\langle \langle \Delta n_{st}^{2}(t)\rangle \rangle _{\{P\}}\equiv \langle
\langle n_{st}^{2}(t)\rangle \rangle _{\{P\}}-\langle \langle
n_{st}(t)\rangle \rangle _{\{P\}}^{2},  \label{Second}
\end{equation}%
can be obtained in a similar way. First, from Eq. (\ref{StochasticIntensity}%
) we write the second moment as%
\begin{eqnarray}
\langle \langle n_{st}^{2}(t)\rangle \rangle _{\{P\}} &=&2\sum_{RR^{\prime
}}\int_{0}^{t}d\tau \int_{0}^{\tau }d\tau ^{\prime }\langle \delta
_{RR_{st}(\tau )}\delta _{R^{\prime }R_{st}(\tau ^{\prime })}\rangle   \notag
\\
&&\frac{d}{d\tau }\frac{d}{d\tau ^{\prime }}[\overline{n_{R}^{st}(\tau
)n_{R^{\prime }}^{st}(\tau ^{\prime })}].  \label{Cuadratic}
\end{eqnarray}%
The average appearing in the first line can be written as 
\begin{subequations}
\begin{eqnarray}
\langle \delta _{RR_{st}(\tau )}\delta _{R^{\prime }R_{st}(\tau ^{\prime
})}\rangle  &=&P(R,\tau ;R^{\prime },\tau ^{\prime }), \\
&=&P(R,\tau |R^{\prime },\tau ^{\prime })P_{R^{\prime }}(\tau ^{\prime }), \\
&=&P(R,\tau |R^{\prime },\tau ^{\prime })P_{R^{\prime }}^{\infty }.
\end{eqnarray}%
By definition, $P(R,\tau ;R^{\prime },\tau ^{\prime })$ is the joint
probability for observing successively the bath in the states $R^{\prime }$
and $R$ at times $\tau ^{\prime }$ and $\tau $\ respectively ($\tau >\tau
^{\prime }$). By using the Markov property of the underlying bath
fluctuations it can be expressed in terms of the conditional probability $%
P(R,\tau |R^{\prime },\tau ^{\prime })$ \cite{vanKampen}. As before, for
simplifying the analysis, in the third line of the previous equation we
assumed that the bath fluctuations begin in their stationary state.

The second line of Eq. (\ref{Cuadratic}) define the correlation of the
counting processes $\{n_{R}^{st}(\tau )\}.$ Due to the statistical
independence of these objects, when $R\neq R^{\prime }$ it follows $%
\overline{n_{R}^{st}(\tau )n_{R^{\prime }}^{st}(\tau ^{\prime })}=\overline{%
n_{R}^{st}}(\tau )\overline{n_{R^{\prime }}^{st}}(\tau ^{\prime
})=I_{R}I_{R^{\prime }}\tau \tau ^{\prime }.$ After some manipulation, the
second cumulant (\ref{Second}) reads 
\end{subequations}
\begin{eqnarray}
\langle \langle \Delta n_{st}^{2}(t)\rangle \rangle _{\!\{P\}}\!\!
&=&\!\!2\sum_{R}\!\!\int_{0}^{t}\!\!d\tau \int_{0}^{\tau }\!\!d\tau ^{\prime
}\frac{d}{d\tau }\frac{d}{d\tau ^{\prime }}\overline{\Delta n_{R}^{2}}(\tau
,\tau ^{\prime })\!\!  \label{integral} \\
&\times &\!\!\!P(R,\tau |R,\tau ^{\prime })P_{R}^{\infty }+2\sum_{RR^{\prime
}}I_{R}I_{R^{\prime }}f_{RR^{\prime }}(t),  \notag
\end{eqnarray}%
where we have introduced the matrix of functions%
\begin{equation}
f_{RR^{\prime }}(t)=\int_{0}^{t}d\tau \int_{0}^{\tau }d\tau ^{\prime
}[P(R,\tau |R^{\prime },\tau ^{\prime })-P_{R}^{\infty }]P_{R^{\prime
}}^{\infty },  \label{ftime}
\end{equation}%
and the diagonal correlation $\overline{\Delta _{R}^{st}}(\tau ,\tau
^{\prime })=\overline{n_{R}^{st}(\tau )n_{R}^{st}(\tau ^{\prime })}-%
\overline{n_{R}^{st}}(\tau )\overline{n_{R}^{st}}(\tau ^{\prime }).$ The
correlation time\ of this object is much smaller than the bath transition
time. Hence, we can approximate $(d/d\tau ^{\prime })\overline{\Delta
n_{R}^{st}}(\tau ,\tau ^{\prime })\approx \delta (\tau -\tau ^{\prime })%
\overline{\Delta n_{R}^{st}}(\tau ,\tau )=\delta (\tau -\tau ^{\prime
})\Delta n_{R}^{st}(\tau ),$ where%
\begin{equation}
\overline{\Delta n_{R}^{2}}(\tau )=\overline{n_{R}^{st}(\tau
)n_{R}^{st}(\tau )}-\overline{n_{R}^{st}}(\tau )\overline{n_{R}^{st}}(\tau ),
\end{equation}%
is the second cumulant of each process $n_{R}^{st}(\tau ).$ From Eq. (\ref%
{integral}) we get the final expression%
\begin{equation}
\langle \langle \Delta n_{st}^{2}(t)\rangle \rangle
_{\{P\}}=\sum_{R}P_{R}^{\infty }\overline{\Delta n_{R}^{2}}%
(t)+2\sum_{RR^{\prime }}I_{R}I_{R^{\prime }}f_{RR^{\prime }}(t).
\label{VarianzaStoca}
\end{equation}%
We notice that in addition to the linear combination given by the first
contribution, $\langle \langle \Delta n_{st}^{2}(t)\rangle \rangle _{\{P\}}$
is also proportional to the intensities $\{I_{R}\}.$ The normalized cumulant 
$\langle \langle \Delta n_{st}^{2}\rangle \rangle =\lim_{t\rightarrow \infty
}(1/t)\langle \langle \Delta n_{st}^{2}(t)\rangle \rangle $ reads%
\begin{equation}
\langle \langle \Delta n_{st}^{2}\rangle \rangle
_{\{P\}}=\sum_{R}P_{R}^{\infty }\overline{\Delta n_{R}^{2}}%
+2\sum_{RR^{\prime }}I_{R}I_{R^{\prime }}f_{RR^{\prime }}.
\label{VarianzaGeneral}
\end{equation}%
The matrix $f_{RR^{\prime }}$ is given by%
\begin{equation}
f_{RR^{\prime }}\equiv \lim_{t\rightarrow \infty }\frac{1}{t}f_{RR^{\prime
}}(t),  \label{festacion}
\end{equation}%
the intensities $\{I_{R}\}$ follows from Eq. (\ref{I_R}) and%
\begin{equation}
\overline{\Delta n_{R}^{2}}=I_{R}\left[ 1-\frac{(6\gamma _{R}^{2}-8\delta
_{R}^{2})\Omega _{R}^{2}}{(\gamma _{R}^{2}+2\Omega _{R}^{2}+4\delta
_{R}^{2})^{2}}\right] ,
\end{equation}%
where as before $\delta _{R}=\omega _{L}-\omega _{R}.$ This expression, as
well as that for the set $\{I_{R}\},$ satisfy the relations 
\begin{equation}
I_{R}=\left. \frac{\partial \Theta _{M}^{\prime }(s)}{\partial s}\right\vert
_{s=0,},\ \ \ \ \overline{\Delta n_{R}^{2}}=-\left. \frac{\partial
^{2}\Theta _{M}^{\prime }(s)}{\partial s^{2}}\right\vert _{s=0,},
\end{equation}%
where $\Theta _{M}^{\prime }$ is the grand potential associated to Eq. (\ref%
{G_Markov}) under the replacements $\omega \rightarrow \omega _{R},$ $\Omega
\rightarrow \Omega _{R},$ and $\gamma \rightarrow \gamma _{R}.$

For a two-dimensional configurational bath space, $R=A,B,$ the constants (%
\ref{festacion}) can easily be written as $f_{RR^{\prime }}=(2\delta
_{RR^{\prime }}-1)P_{A}^{\infty }P_{B}^{\infty }.$ From Eq. (\ref%
{VarianzaGeneral}), we get%
\begin{equation}
\langle \langle \Delta n_{st}^{2}\rangle \rangle
_{\{P\}}=\sum_{R=A,B}P_{R}^{\infty }\overline{\Delta n_{R}^{2}}+2\frac{%
P_{A}^{\infty }P_{B}^{\infty }(I_{A}-I_{B})^{2}}{(\phi _{AB}+\phi _{BA})},
\label{VarianzaTwo}
\end{equation}%
where $P_{A}^{\infty }=\phi _{AB}/(\phi _{AB}+\phi _{BA})$ and $%
P_{B}^{\infty }=\phi _{BA}/(\phi _{AB}+\phi _{BA}).$ We notice that these
expressions recover the results obtained in Ref. \cite{slow} through a
different approach.

\subsection{Extension to the thermodynamic approach}

The previous stochastic approach describes the $s$-ensemble in $s=0,$ i.e.,
in a slow modulation limit it can fit the statistical behavior dictated by
the probabilities $\{P_{n}(t)\}$ [Eq. (\ref{Pn})] or equivalently the states 
$\{\rho _{R}(t)\}$ [Eq. (\ref{Rho_S})]. For $s\neq 0,$ it does not apply. In
this case, the relevant objects are the probabilities $\{q_{n}(t)\}$ [Eq. (%
\ref{q})] and the states $\{\tilde{\rho}_{R}(t)\}$ [Eq. (\ref{RhoTilde})].
In the following calculations, we assume that the slow modulation limit and
the stochastic approach can also be applied for any value of the pseudo
chemical potential $s.$ The consistency of this ansatz relies on the results
that can be obtained from it.

By noting that $\langle \langle n_{st}\rangle \rangle _{\{P\}}=\langle
\langle N\rangle \rangle |_{s=0},$ from Eq. (\ref{medioStoca}), we write the
first $s$-dependent cumulant as%
\begin{equation}
\langle \langle N\rangle \rangle \simeq \sum_{R}\tilde{P}_{R}^{\infty
}(s)\langle \langle N_{R}\rangle \rangle ,  \label{ENEs}
\end{equation}%
where consistently $\langle \langle N_{R}\rangle \rangle \equiv \langle
\langle N\rangle \rangle _{M}^{\prime }.$ Here, $\langle \langle N\rangle
\rangle _{M}^{\prime }$ is defined by Eq. (\ref{CumulanteMarkov}) after
replacing $\omega \rightarrow \omega _{R},$ $\Omega \rightarrow \Omega _{R},$
and $\gamma \rightarrow \gamma _{R}.$ On the other hand, here the weights are%
\begin{equation}
\tilde{P}_{R}^{\infty }(s)\equiv \lim_{t\rightarrow \infty }\mathrm{Tr}[%
\tilde{\rho}_{R}(t)],  \label{EstacionS}
\end{equation}%
where $\tilde{\rho}_{R}(t)$ is defined by (\ref{RhoTilde}).

Taking into account that $\langle \langle \Delta n_{st}^{2}\rangle \rangle
_{\{P\}}=\langle \langle \Delta N^{2}\rangle \rangle |_{s=0},$ from Eq. (\ref%
{VarianzaStoca}) we propose the expression%
\begin{eqnarray}
\langle \langle \Delta N^{2}\rangle \rangle &\simeq &\sum_{R}\tilde{P}%
_{R}^{\infty }(s)\langle \langle \Delta N_{R}^{2}\rangle \rangle
\label{DeltaNs} \\
&&+2\sum_{RR^{\prime }}\langle \langle N_{R}\rangle \rangle \langle \langle
N_{R^{\prime }}\rangle \rangle \tilde{f}_{RR^{\prime }}(s),  \notag
\end{eqnarray}%
where $\langle \langle \Delta N_{R}^{2}\rangle \rangle \equiv \langle
\langle \Delta N^{2}\rangle \rangle _{M}^{\prime }$ also follows from Eq. (%
\ref{CumulanteMarkov}) after replacing $\omega \rightarrow \omega _{R},$ $%
\Omega \rightarrow \Omega _{R},$ and $\gamma \rightarrow \gamma _{R}.$

The expressions (\ref{ENEs})\ and (\ref{DeltaNs}) depend on the stationary
populations (\ref{EstacionS}) and the generalized matrix $\tilde{f}%
_{RR^{\prime }}(s).$ Its definition can read from Eq. (\ref{festacion})
after introducing in Eq. (\ref{ftime}) the $s$-dependence of the bath
populations [$P_{R}\rightarrow \tilde{P}_{R}(s)$]. The exact expressions for
these objects for arbitrary bath spaces [Eq. (\ref{Classical})] are very
complicated and do not provide an intuitive frame for understanding the
thermodynamics of the $s$-ensemble. Therefore, from now on we restrict to a
two-dimensional reservoir, $R=A,B.$ From Eq. (\ref{VarianzaTwo}) we write%
\begin{eqnarray}
\langle \langle \Delta N^{2}\rangle \rangle &\simeq &\sum_{R=A,B}\tilde{P}%
_{R}^{\infty }(s)\langle \langle \Delta N_{R}^{2}\rangle \rangle +2\frac{%
\tilde{P}_{A}^{\infty }(s)\tilde{P}_{B}^{\infty }(s)}{\tilde{\phi}(s)} 
\notag \\
&&\times (\langle \langle N_{A}\rangle \rangle -\langle \langle N_{B}\rangle
\rangle )^{2},  \label{DeltaNsTWO_R}
\end{eqnarray}%
where $\tilde{\phi}(s)=\tilde{\phi}_{AB}(s)+\tilde{\phi}_{BA}(s).$

Eqs. (\ref{ENEs}) and (\ref{DeltaNsTWO_R}) only depends on the stationary
populations $\{\tilde{P}_{R}^{\infty }(s)\}$ and the generalized rate $%
\tilde{\phi}(s).$ These functions can be approximated in the following way.
Consistently with an equilibrium thermodynamic approach, the populations are
expressed as%
\begin{equation}
\tilde{P}_{A}^{\infty }(s)\simeq \frac{e^{-[\zeta _{A}+\xi _{A}(s)]}}{%
\mathcal{Z}(s)},\ \ \ \ \ \ \ \tilde{P}_{B}^{\infty }(s)\simeq \frac{%
e^{-[\zeta _{B}+\xi _{B}(s)]}}{\mathcal{Z}(s)},  \label{PopExponential}
\end{equation}%
where $\zeta _{R}+\xi _{R}(s)$ is the \textquotedblleft
energy\textquotedblright\ associated to each $R$-bath state. This splitting
is defined such that $\xi _{R}(0)=0.$ On the other hand, the function $%
\mathcal{Z}(s)$ guarantees the normalization $\tilde{P}_{A}^{\infty }(s)+%
\tilde{P}_{B}^{\infty }(s)=1.$ Then, we can rewrite 
\begin{subequations}
\label{popuTANH}
\begin{eqnarray}
\tilde{P}_{A}^{\infty }(s) &\simeq &\frac{1}{2}\{1-\tanh [\varepsilon
_{0}+\varepsilon (s)]\}, \\
\tilde{P}_{B}^{\infty }(s) &\simeq &\frac{1}{2}\{1+\tanh [\varepsilon
_{0}+\varepsilon (s)]\},
\end{eqnarray}%
where $\varepsilon _{0}=(\zeta _{A}-\zeta _{B})/2,$ and $\varepsilon
(s)=[\xi _{A}(s)-\xi _{B}(s)]/2.$ In $s=0,$ these expressions must to
satisfy the condition $\tilde{P}_{R}^{\infty }(0)=P_{R}^{\infty },$ which
imply the expression 
\end{subequations}
\begin{equation}
\varepsilon _{0}=\frac{1}{2}\log \left[ \frac{P_{B}^{\infty }}{P_{A}^{\infty
}}\right] =\frac{1}{2}\log \left[ \frac{\phi _{BA}}{\phi _{AB}}\right] .
\label{energia}
\end{equation}%
On the other hand, the dependence of $\varepsilon (s)$ is assumed to be
linear in $s,$ i.e., $\varepsilon (s)\simeq s\delta \varepsilon .$ The
constant $\delta \varepsilon $ can be determine from the relation between
Eqs. (\ref{ENEs}) and (\ref{DeltaNs}) in $s=0,$ i.e., $(\partial /\partial
s)\langle \langle N\rangle \rangle |_{s=0}=-\langle \langle \Delta
N^{2}\rangle \rangle |_{s=0}.$ After some algebra, from Eq. (\ref%
{DeltaNsTWO_R}), it follows%
\begin{equation}
\varepsilon (s)\simeq s\ \delta \varepsilon =s\ \frac{I_{A}-I_{B}}{\phi
_{AB}+\phi _{BA}}.  \label{chemical}
\end{equation}%
Both $\varepsilon _{0}$ and $\delta \varepsilon $ may assume positive and
negative values.

The generalized rate $\tilde{\phi}(s)$ [Eq. (\ref{DeltaNsTWO_R})] defines
the characteristic decay time of the $s$-dependent bath populations. It must
to satisfies the conditions $\tilde{\phi}(s)\geq 0$ and $\tilde{\phi}%
(0)=(\phi _{AB}+\phi _{BA}).$ We assume the dependence%
\begin{equation}
\tilde{\phi}(s)\simeq \alpha s(s-2s_{p})+(\phi _{AB}+\phi _{BA}),
\label{rateS}
\end{equation}%
where $s_{p}$ is a dimensionless parameter while $\alpha \geq 0,$ as well as 
$\tilde{\phi}(s),$ has units of [1/time]. As a function of $s,$ the rate $%
\tilde{\phi}(s)$ reaches its minimal value at $s=s_{p}.$ This constant is
chosen as the value of $s$ at which the second contribution of Eq. (\ref%
{DeltaNsTWO_R}) reaches its maximal value. It can be approximated by the
value at which the function $\tilde{P}_{A}^{\infty }(s)\tilde{P}_{B}^{\infty
}(s)=1/(2\cosh [\varepsilon _{0}+\varepsilon (s)])^{2}$ is maximal. Hence,
we take%
\begin{equation}
s_{p}=-\frac{\varepsilon _{0}}{\delta \varepsilon }=\frac{(\phi _{AB}+\phi
_{BA})}{2(I_{A}-I_{B})}\log \left[ \frac{P_{A}^{\infty }}{P_{B}^{\infty }}%
\right] .  \label{sPico}
\end{equation}%
Finally, the coefficient $\alpha $ can be determine from the condition $%
(\partial /\partial s)\langle \langle \Delta N^{2}\rangle \rangle
|_{s=0}=-\langle \langle \Delta N^{3}\rangle \rangle |_{s=0}.$
Alternatively, due to its large size expression, it can be considered as a
free fitting parameter.

The $s$-extension of the stochastic approach allows us to get closed
expressions [Eqs. (\ref{ENEs}) and (\ref{DeltaNsTWO_R})] for approximating
the thermodynamic approach in the slow modulation limit. In Fig. 2 we plot
the average number and the normalized fluctuations obtained from Eqs. (\ref%
{Zasimptotico}), (\ref{ZfromG}) and (\ref{G_R}).%
\begin{figure}[tb]
\includegraphics[bb=34 34 435 624,angle=0,width=8.7cm]{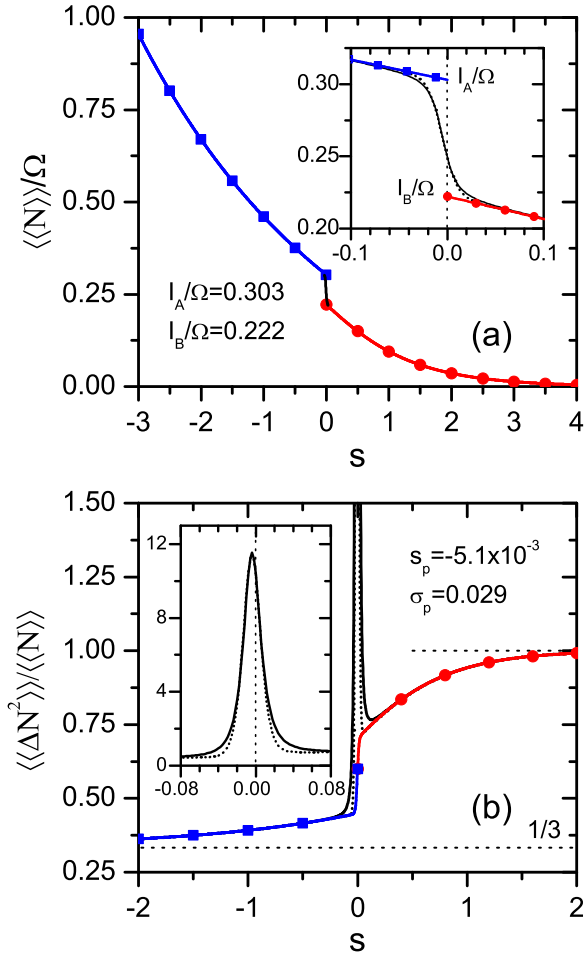}
\caption{(color online) Average number value $\langle \langle N\rangle
\rangle $ and normalized fluctuations $\langle \langle \Delta N^{2}\rangle
\rangle /\langle \langle N\rangle \rangle $ as a function of $s$ in the slow
modulation limit. The insets show the behaviors around $s\approx 0.$ The
full (black) curves correspond to the numerical solutions associated to Eqs.
(\protect\ref{Zasimptotico}), (\protect\ref{ZfromG}), and (\protect\ref{G_R}%
). The dotted (black) curves are the fitting obtained from the stochastic
approach, Eqs. (\protect\ref{ENEs}) and (\protect\ref{DeltaNsTWO_R}). In (a)
the blue squares and red circles correspond respectively to $\langle \langle
N_{A}\rangle \rangle $ and $\langle \langle N_{B}\rangle \rangle $
determined from Eqs. (\protect\ref{CumulanteMarkov}) and (\protect\ref%
{GranMarkov}). In (b) the blue squares and red circles correspond to $%
\langle \langle \Delta N_{A}^{2}\rangle \rangle /\langle \langle N\rangle
\rangle $ and $\langle \langle \Delta N_{B}^{2}\rangle \rangle /\langle
\langle N\rangle \rangle .$ The parameters are $\protect\omega _{R}=\protect%
\omega _{0},$ $\Omega _{R}=\Omega ,$ $\protect\omega _{L}=\protect\omega %
_{0},$ $\protect\gamma _{A}/\Omega =2.5,$ $\protect\gamma _{B}/\Omega =0.5,$ 
$\protect\phi _{AB}/\Omega =4\times 10^{-4},$ $\protect\phi _{BA}/\Omega
=8\times 10^{-4},$ and $\protect\alpha /\Omega =2.15.$}
\end{figure}
As in the previous section, the configurational bath space is
two-dimensional, $R=A,B.$ It only affects the natural decay of the system, $%
\{\gamma _{R}\}.$ Hence, we take $\omega _{R}=\omega _{0},$ $\Omega
_{R}=\Omega ,$ and $\omega _{L}=\omega _{0},$ i.e., the spectral shifts are
null, the system-laser interaction is independent of the bath states, and
the laser excitation is in resonance with the system.

In Fig. 2(a) we plot $\langle \langle N\rangle \rangle $ as a function of $%
s. $ The full (black) curve is the (exact) numerical solution obtained from
Eqs. (\ref{Zasimptotico}), (\ref{ZfromG}), and (\ref{G_R}). The grand
potential $\Theta (s)$ follows from the larger root of an eighth order
polynomial function. The dotted (black) curves (indistinguishable)
correspond to the fitting arising from the stochastic approach, Eq. (\ref%
{ENEs}). In the scale of the plot, we note that the average number can be
approximated as%
\begin{equation}
\langle \langle N\rangle \rangle \approx \left\{ 
\begin{array}{c}
\langle \langle N_{A}\rangle \rangle \ \ \ \ for\ s<0 \\ 
\langle \langle N_{B}\rangle \rangle \ \ \ \ for\ s>0%
\end{array}%
\right. ,\ \ \ \ (I_{A}>I_{B}).  \label{Naprox}
\end{equation}%
The superposed blue squares and red circles curves correspond to $\langle
\langle N_{A}\rangle \rangle $ and $\langle \langle N_{B}\rangle \rangle $
respectively. These contributions follow as the first derivative with
respect to $s$ of the grand potential $\Theta _{M}(s),$ Eq. (\ref{GranMarkov}%
), under the replacement $\gamma \rightarrow \gamma _{R}.$ The (crude)
approximation (\ref{Naprox}) implies that [see Eq. (\ref{ENEs})] $\tilde{P}%
_{A}^{\infty }(s)\approx \theta (-s)$ and $\tilde{P}_{B}^{\infty }(s)\approx
\theta (s),$ where $\theta (s)$ is the step function $[\theta (s)=0$ for $%
s<0 $ and $\theta (s)=1$ for $s>0].$ As the values of the decay rates $%
[\gamma _{A}/\Omega =2.5,$ $\gamma _{B}/\Omega =0.5]$ correspond to the
average values of Fig. 1, the behavior of $\langle \langle N_{A}\rangle
\rangle $ and $\langle \langle N_{B}\rangle \rangle $ for any value of $s$
can also be read from that plot.

Consistently with the approximation (\ref{Naprox}), the behavior of $\langle
\langle N\rangle \rangle $ around $s\approx 0$ seems to be discontinuous. We
notice that a similar behavior arises in thermodynamical first-order
transitions. For example, if a transition is driven by temperature, the
discontinuity in the derivative of the thermodynamic potential may
corresponds to the difference of specific volume of two coexisting phases 
\cite{raquel}. Here, the \textquotedblleft jump\textquotedblright\ in $%
\langle \langle N\rangle \rangle $ is $(\langle \langle N_{A}\rangle \rangle
-\langle \langle N_{B}\rangle \rangle )|_{s=0}=I_{A}-I_{B}.$ Hence, two
thermodynamic phases can be associated to the intensity regimes defined by $%
\langle \langle N_{A}\rangle \rangle $ and $\langle \langle N_{B}\rangle
\rangle .$

While the average $\langle \langle N\rangle \rangle $ seems to be a
discontinuous function on a large $s$-scale, around the origin it is a
continuous function of $s.$ This property is shown in the inset of Fig.
2(a). Even at those small scales, the stochastic approach, Eq. (\ref{ENEs}),
provides an indistinguishable fitting (black dotted curve). In
thermodynamical systems, finite-size effects produce a similar smoothing of
the free energy derivative \cite{binderBook,binder,landauIsing,challa}. In
the present case, this relation is established in the following section.

In Fig. 2(b) we plot the normalized fluctuations $\langle \langle \Delta
N^{2}\rangle \rangle /\langle \langle N\rangle \rangle $ as a function of $%
s. $ Consistently with the rough approximation (\ref{Naprox}), on larger $s$%
-scales we expect the validity of the approximation%
\begin{equation}
\langle \langle \Delta N^{2}\rangle \rangle \approx \left\{ \!\!%
\begin{array}{c}
\langle \langle \Delta N_{A}^{2}\rangle \rangle \ \ \ \ for\ s<0 \\ 
\langle \langle \Delta N_{B}^{2}\rangle \rangle \ \ \ \ for\ s>0%
\end{array}%
\right. ,\ \ \ \ (I_{A}>I_{B}),  \label{FluctBrut}
\end{equation}%
where $\langle \langle \Delta N_{R}^{2}\rangle \rangle $ follows from Eqs. (%
\ref{CumulanteMarkov}) and (\ref{GranMarkov}). The superposed blue squares
and red circles curves correspond to this approximation. While they provide
a very good fitting for $|s|>0,$ around the origin the fluctuations develops
a narrow and abrupt peak (see the inset). The stochastic approach (black
dotted line) also fits this property.

The background behavior and the peak in $\langle \langle \Delta N^{2}\rangle
\rangle $ can be read from Eq. (\ref{DeltaNsTWO_R}). In fact, the crude
approximation (\ref{FluctBrut}) is indistinguishable from the contribution $%
\sum_{R=A,B}\tilde{P}_{R}^{\infty }(s)\langle \langle \Delta
N_{R}^{2}\rangle \rangle .$ On the other hand, the narrow peak is fitted by
the contribution proportional to the product of the stationary populations $%
\tilde{P}_{A}^{\infty }(s)\tilde{P}_{B}^{\infty }(s)=1/(2\cosh [\varepsilon
_{0}+\varepsilon (s)])^{2}.$ Therefore, the maximal value of the peak occurs
at $s=s_{p},$ Eq. (\ref{sPico}), i.e., the value of $s$ at which $\tilde{P}%
_{A}^{\infty }(s)=\tilde{P}_{B}^{\infty }(s)=1/2.$ The inset of Fig. 2(b)
confirms this prediction. Furthermore, from Eq. (\ref{DeltaNsTWO_R}) the
value of $\langle \langle \Delta N^{2}\rangle \rangle $ at $s_{p}$ can be
approximated as%
\begin{equation}
\langle \langle \Delta N^{2}\rangle \rangle |_{s=s_{p}}\approx \frac{1}{2}%
\sum_{R=A,B}\overline{\Delta n_{R}^{2}}+\frac{(I_{A}-I_{B})^{2}}{2(\phi
_{AB}+\phi _{BA})},  \label{ValorPico}
\end{equation}%
$[\langle \langle N\rangle \rangle |_{s=s_{p}}\approx (I_{A}+I_{B})/2],$
while the width of the peak, $\sigma _{p},$ can be estimated as%
\begin{equation}
\sigma _{p}\approx 2\frac{(\phi _{AB}+\phi _{BA})}{|I_{A}-I_{B}|}.
\label{AnchoPico}
\end{equation}

Taking different values of the parameters of the evolution (\ref{G_R}), we
have checked that in the slow modulation limit the position, height and
width of the peak obey the scaling defined by Eqs. (\ref{sPico}), (\ref%
{ValorPico}), and (\ref{AnchoPico}) respectively. From these expressions,
one can deduce that in the limit $(\phi _{AB}+\phi _{BA})\rightarrow 0,$ the
peak becomes proportional to a delta Dirac function. Therefore,
asymptotically a first-order transition happens. The thermodynamic response
functions, for all values of $s,$ are given by Eqs. (\ref{Naprox}) and (\ref%
{FluctBrut}), i.e., the grand potential is $\Theta (s)=\Theta _{A}(s)$ for $%
s<0,$ $\Theta (s)=\Theta _{B}(s)$ for $s>0,$ with $\Theta (0)=0.$

\subsection{Finite-size effects and double-Gaussian approximation}

Finite-size effects in first-order transitions \cite{binderBook,binder} has
been analyzed for systems such as the Ising \cite{landauIsing} and q-state
Pott models \cite{challa}. While in these systems the transition is driven
by a magnetic field or temperature, the thermodynamic functions have similar
behaviors to those shown in Fig 2. The scaling of the peak in the second
derivative of the thermodynamic potential \cite{challa} is similar to those
of Eqs. (\ref{sPico}), (\ref{ValorPico}), and (\ref{AnchoPico}).

From a theoretical point of view, finite-size effects at first-order
transitions can be characterized over the basis of (equilibrium) Einstein
fluctuation theory \cite{raquel}, which provides the probability
distribution of the thermodynamic variable fluctuations. For example, for an
open (thermal) system, the probability distribution $\mathcal{P}(N)$ of the
particle number is a Gaussian distribution $\mathcal{P}(N)=[2\pi kT(\partial
/\partial \mu )\bar{N}]^{-1/2}\exp [-(N-\bar{N})^{2}/2kT(\partial /\partial
\mu )\bar{N}],$ where $T$ is the temperature, $\bar{N}$ is the average
particle number and $\mu $ is the chemical potential.

In the present approach, the transition is driven by the pseudo chemical
potential $s$ and the size of the system must to be inversely proportional
to the rate of the bath fluctuations. Consistently with the Einstein
fluctuation theory, we search for a probability distribution $\mathcal{P}%
(N), $ with $\int_{-\infty }^{+\infty }\mathcal{P}(N)dN=1,$ such that the
average number can be obtained as%
\begin{equation}
\langle \langle N\rangle \rangle =\int_{-\infty }^{+\infty }\mathcal{P}%
(N)NdN,  \label{AverageGauss}
\end{equation}%
while the second cumulant follows from%
\begin{equation}
\langle \langle \Delta N^{2}\rangle \rangle =\frac{1}{\Upsilon (s)}%
\int_{-\infty }^{+\infty }\mathcal{P}(N)(N-\langle \langle N\rangle \rangle
)^{2}dN.  \label{FluctGauss}
\end{equation}%
Here, the inverse of $\Upsilon (s)$\ measures the \textquotedblleft
size\textquotedblright\ of the system \cite{challa}. In an infinite-size
limit, $\mathcal{P}(N)$ must be a Gaussian distribution \cite{raquel}.
Nevertheless, when finite-size effects are considered in a first-order
transition, one must to consider a superposition of Gaussian distributions,
each one representing the coexisting phases \cite{challa}. In fact, the
different phases are randomly explored by the system when its size is finite 
\cite{challa}. This effect is similar to the blinking property of the slow
modulation limit \cite{SMS,SMSJumps}.%
%
\begin{figure}[tb]
\includegraphics[bb=50 331 440 624,angle=0,width=8.7cm]{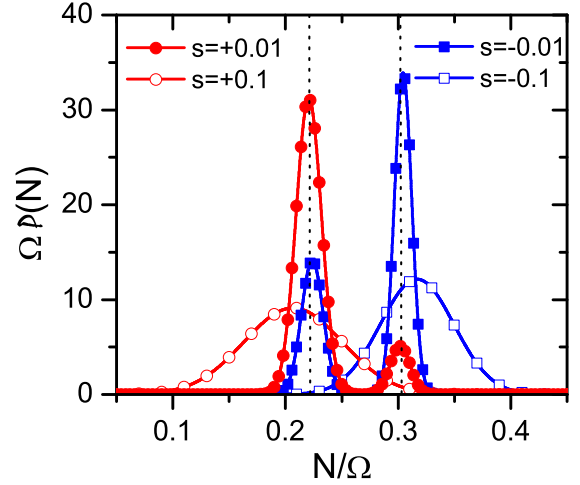}
\caption{(color online) Probability distribution (\protect\ref{Gauss}) for
different values of $s.$ The vertical dotted lines correspond to $N/\Omega
=I_{A}/\Omega =0.303$, and $N/\Omega =I_{B}/\Omega =0.222.$ The parameters
are the same than in Fig. 2.}
\end{figure}

In our problem, the coexistent phases correspond to the different intensity
regimes defined by $\langle \langle N_{R}\rangle \rangle .$ Then,%
\begin{eqnarray}
\mathcal{P}(N)\!\! &=&\!\!\frac{\tilde{P}_{A}^{\infty }(s)}{\sqrt{2\pi
\langle \langle \Delta N_{A}^{2}\rangle \rangle \Upsilon (s)}}\exp \!\!\left[
\!-\frac{(N-\langle \langle N_{A}\rangle \rangle )^{2}}{2\langle \langle
\Delta N_{A}^{2}\rangle \rangle \Upsilon (s)}\!\right]   \label{Gauss} \\
&&\!\!\!+\frac{\tilde{P}_{B}^{\infty }(s)}{\sqrt{2\pi \langle \langle \Delta
N_{B}^{2}\rangle \rangle \Upsilon (s)}}\exp \!\!\left[ \!-\frac{(N-\langle
\langle N_{B}\rangle \rangle )^{2}}{2\langle \langle \Delta N_{B}^{2}\rangle
\rangle \Upsilon (s)}\!\right] .  \notag
\end{eqnarray}%
Here, $\langle \langle N_{R}\rangle \rangle $ and $\langle \langle \Delta
N_{R}^{2}\rangle \rangle $ follows from Eq. (\ref{CumulanteMarkov}) while $%
\Upsilon (s)=\tilde{\phi}(s)/2$ [Eq. (\ref{rateS})]. Furthermore, the
weight\ of each Gaussian distribution is expressed in terms of the
thermodynamic potential of each phase \cite{challa}. Therefore, the
populations $\tilde{P}_{R}^{\infty }(s)$ are written as $\tilde{P}%
_{R}^{\infty }(s)\simeq \exp [-\zeta _{R}-\Theta _{R}(s)/\tilde{\phi}(s)]/%
\mathcal{Z}(s),$ where $\mathcal{Z}(s)$ guarantees the normalization $\tilde{%
P}_{A}^{\infty }(s)+\tilde{P}_{B}^{\infty }(s)=1$ [compare with (\ref%
{PopExponential})] and $\Theta _{R}(s)$ is the grand potential of each
phase. For the example shown in Fig. 2, these functions can be read from Eq.
(\ref{GranMarkov}) under the replacement $\gamma \rightarrow \gamma _{R}.$
Trivially, the populations $\tilde{P}_{R}^{\infty }(s)$ can be rewritten as
in Eq. (\ref{popuTANH}) with $\varepsilon _{0}$ defined by Eq. (\ref{energia}%
) and%
\begin{equation}
\varepsilon (s)\approx \frac{\Theta _{A}(s)-\Theta _{B}(s)}{\tilde{\phi}(s)}%
\approx s\frac{I_{A}-I_{B}}{(\phi _{AB}+\phi _{BA})}+O(s^{2}).
\end{equation}%
This result recovers Eq. (\ref{chemical}) and proofs the consistency of the
previous results and scaling. In fact, the double-Gaussian probability
distribution (\ref{Gauss}), through Eqs. (\ref{AverageGauss}) and (\ref%
{FluctGauss}), recovers the expressions of the extended stochastic approach,
Eqs. (\ref{ENEs}) and (\ref{DeltaNsTWO_R}) respectively.

In Fig. 3 we plot the distribution (\ref{Gauss}) for different values of $s.$
The parameters of the underlying evolution are the same than in Fig. 2. Near
of the transition, $s\approx s_{p}\approx 0,$ the probability distribution
is a double-Gaussian one. Consistently, the higher peaks are centered around
the values $N\approx I_{A}$ $(s<0)$ and $N\approx I_{B}$ $(s>0).$ For $%
|s|\gg s_{p},$ the distribution has only one peak, which is centered around
\ $s\approx \langle \langle N_{R}\rangle \rangle .$

The double-Gaussian approximation, and consistently the stochastic approach,
can be extended beyond the slow modulation limit. Nevertheless, parameters
such as the position $s_{p}$ and width $\sigma _{p}\ $of the peak must be
taken as free parameters. For example, a reasonable fitting is obtained
after replacing the polynomial function (\ref{rateS}) with an hyperbolic one
and introducing a non-linear function $\varepsilon (s),$ both of them
defined with extra free parameters.

\section{Summary and Conclusions}

The (non-equilibrium) ensemble of measurement realizations of an open
quantum system can be analyzed with the LD formalism. For fluorescent
systems under a direct photon detection scheme, this approach allow to
describe the asymptotic behavior of the photon counting probabilities
through a thermodynamic-like formalism \cite{garrahan}. In this paper we
have studied the thermodynamic approach associated to a fluorescent system
coupled to a complex self-fluctuating environment able to modify the
characteristic parameters of the system evolution.

The statistical mechanics underlying the thermodynamic frame is defined by a
set of auxiliary probabilities whose characteristic events are the unlikely
ones of the photon counting realizations, Eq. (\ref{q}). A free energy
function, Eq. (\ref{Grand}), through a Legendre transformation, defines the
long time behavior of the photon counting probabilities. Here, its
functional form follows from the trace of a generating function operator,
Eq. (\ref{ZfromG}), whose evolution, Eq. (\ref{G_R}), takes into account the
parameter fluctuations induced by the environment.

In a fast modulation limit, i.e., when the characteristic time of the
environment fluctuations is much smaller than the time between consecutive
photon emissions, the thermodynamic frame can be well approximated with that
corresponding to a Markovian fluorescent system. Its evolution is defined by
a set of parameters that follows from an average weighted by the stationary
populations of each bath state, Eq. (\ref{averageParameters}). When the bath
only affects the natural decay of the system and the external laser
excitation is in resonance with the system, the thermodynamic potential can
be approximated by a simple analytical expression, Eq. (\ref{GranMarkov}).
In this limit, the response functions do not display any property related to
a phase transition. Nevertheless, the normalized fluctuations always
interpolate between a scale invariant regime and a Poissonian one [Fig. 1].

In the slow modulation limit, the fluorescent signal is characterized by a
blinking phenomenon. The scattered intensity randomly changes between a set
of values associated to each bath state, Eq. (\ref{I_R}). The photon
counting process can be approximated by the product of two kind of
statistically independent stochastic variables, one related to the quantum
photon emission process and the other to the bath fluctuations, Eq. (\ref%
{StochasticIntensity}). In the thermodynamic frame, each intensity regime
can be read as a different thermodynamic phase. A natural extension of the
stochastic approach provides the basis for characterizing its statistical
properties. After imposing some consistency relations, the average number is
written as a linear combination of the values corresponding to each phase,
Eq. (\ref{ENEs}). The fluctuations around the average number can be
approximated in a similar way, Eq. (\ref{DeltaNsTWO_R}).

The behavior of the average number and the centered fluctuations is similar
to that found in finite-size systems near a first-order phase transition.
This is the main result of this contribution. Instead of a discontinuity in
the first derivative of the thermodynamic potential, an abrupt but
continuous change in its slope is observed. Furthermore, the second
derivative, instead of a delta Dirac contribution, displays a narrow peak
[Fig.~2]. These effects are controlled by the size of the system, which is
proportional to the transition rate between the bath states. The location,
height and width of the peak obey the scaling properties obtained from the
stochastic approach, i.e., Eqs. (\ref{sPico}), (\ref{ValorPico}), and (\ref%
{AnchoPico}) respectively. The finite-size effects can also be obtained from
a generalization of the Einstein's fluctuations theory. The probability
distribution of the fluctuations follows from a double-Gaussian distribution
[Eq. (\ref{Gauss})], each contribution being related to each coexisting
phase [Fig. 3].

From our results, we conclude that whenever a (photon) counting process has
an underlying blinking property, the thermodynamic approach is characterized
by finite-size effects corresponding to a first-order transition. Therefore,
the studied phenomena, for example, must also appear when the blinking
properties depend on the external laser excitation, i.e., for light assisted
processes \cite{SMS,SMSJumps}. These and previous results \cite{garrahan}
confirm that diverse thermodynamical properties of many body (equilibrium)
systems are also present in the statistical properties of (non-equilibrium)
quantum measurement trajectories. This mapping raise up fundamental physical
questions such as the possibility of simulating complex dynamics with simple
open quantum systems subjected to a continuous measurement process.

\section*{Acknowledgments}

The author thanks fruitful discussions with M. Fiori, E. Urdapilleta, and L.
Quiroga Puello. This work was supported by CONICET, Argentina, PIP
11420090100211.

\end{document}